\documentclass[twocolumn,aps,showpacs,prb,superscriptaddress,tightenlines,amsmath,amssymb]{revtex4}
\usepackage{graphicx}
\usepackage{amssymb}
\usepackage{dcolumn}
\usepackage{amsmath}
\usepackage{bm}% bold math

\usepackage{colordvi}

\begin{document}

\title{Hole spin relaxation in $p$-type GaAs quantum wires investigated
     by numerically solving fully microscopic kinetic spin Bloch
     equations}

\author{C. L\"u}
\affiliation{Hefei National Laboratory for Physical Sciences at
Microscale, University of Science and Technology of China, Hefei,
Anhui, 230026, China}
\affiliation{Department of Physics, University
of Science and Technology of China, Hefei, Anhui, 230026, China}
\author{U. Z\"ulicke}
\affiliation{Institute of Fundamental Sciences and MacDiarmid
Institute for Advanced Materials and Nanotechnology, Massey
University, Private Bag 11~222, Palmerston North, New Zealand}
\author{M.\ W.\ Wu}
\thanks{Author to  whom correspondence should be addressed}
\email{mwwu@ustc.edu.cn.}
\affiliation{Hefei National Laboratory for Physical Sciences at
Microscale, University of Science and Technology of China, Hefei,
Anhui, 230026, China}
\affiliation{Department of Physics, University of Science and Technology
of China, Hefei, Anhui, 230026, China}
\date{\today}

\begin{abstract}
  We investigate the spin relaxation of $p$-type GaAs
  quantum wires by numerically solving the fully microscopic
  kinetic spin Bloch equations. We find that the quantum-wire size
  influences the spin relaxation time effectively by modulating
  the energy spectrum and the heavy-hole--light-hole mixing of
  wire states. The effects of quantum-wire size, temperature,
  hole density, and initial
  polarization are investigated in detail. We show that, depending
  on the situation, the spin relaxation time can either increase or
  decrease with hole density. Due to the different subband structure
  and effects arising from spin-orbit coupling, many spin-relaxation
  properties are quite different from those of holes in the bulk or in
  quantum wells, and the inter-subband scattering makes a marked
  contribution to the spin relaxation.
\end{abstract}
\pacs{72.25.Rb, 73.21.Hb, 71.10.-w}
\maketitle

\section{Introduction}

Spintronics is continuing to attract considerable interest because of its
potential application to information technology.\cite{spintronics}
Several spintronics devices have been proposed that manipulate spin via
spin-orbit coupling (SOC).\cite{datta,Loss,filter} In recent years, progress in
nanofabrication and growth techniques has made it possible to produce
high-quality quantum wires (QWRs) and investigate spin physics in
semiconductor nanostructures.\cite{Quay, Ritchie, Hirayama} The energy
spectrum of QWR systems with strong SOC has been extensively
studied.\cite{Cahay2, Liu, Ulrich1,Ulrich2,Hausler,Arakawa, Bastard,
Kapon, Fasolino,Chang,Bimberg,Vlaev} It is well known
that, even in the absence of an external magnetic field, the Rashba\cite{Rashba}
 and Dresselhaus\cite{Dresselhaus}
 SOCs lift the spin degeneracy in wire subbands at nonzero
wave vectors. The subband structure for quantum-confined valence-band
states is particularly interesting since they are subject to an especially strong
SOC.\cite{Ulrich1,Ulrich2,Arakawa,Bastard,Chang,Kapon,Fasolino,Sherman,Bimberg,Vlaev}

As a long spin relaxation time (SRT) is desirable for the operation of spintronic
devices, many investigations have been performed to better understand  the
electron spin relaxation in quantum
structures,\cite{Bleibaum,Cahay1, Ridley, Raimondi, Ando, Rossler, Flatte, Bottesi,Privman,Yoh,Smith,Balents,Wu1,Wu2, Wu3, Weng, Wu5, review, Cheng, schu,Zhou, Zhou2, Cheng2,Clv} 
e.g., using the single-particle
model\cite{Rossler,Flatte,Bleibaum,Cahay1,Ridley,Raimondi,Ando} or Monte-Carlo 
simulations.\cite{Privman,Yoh,Cahay1} However, it was shown by Wu {\em et
al.}\cite{Wu1,Wu2,Wu3,Weng,Wu5} that the single-particle approach is
inadequate in accounting for the spin relaxation. A fully microscopic
kinetic spin Bloch equation (KSBE) theory,
which takes full account of the inhomogeneous broadening
from the Dresselhaus and/or Rashba SOC and the effect of
 scattering, has been developed  to study spin
relaxation.\cite{review,Wu1,Wu2,Wu3,Weng,Wu5} Cheng {\em et al.}
applied this approach (excluding the Coulomb scattering) to study electron spin
relaxation in QWR systems and showed the feasibility of manipulating
spin decoherence.\cite{Cheng} Investigations of spin relaxation of {\em holes\/}
in QWRs are relatively limited,\cite{Ando} even though knowledge of hole spin
relaxation in $p$-type QWRs is important for assessing
the feasibility of hole-based
spintronic devices.\cite{ohno}
The spin-relaxation mechanism in hole QWRs can be expected to be quite different
from that in electron systems, and 2D or bulk hole systems, due to the strong
SOC and the complex wire-subband structure. These effects have not been
addressed previously.

In this paper, we investigate hole spin relaxation in a 
$p$-doped (001) GaAs QWR. An idealized system of quantum
wire with rectangular confinements and hard wall potential is
considered in our calculation. First, we obtain the subband
structure by diagonalizing the hole Hamiltonian including the quantum
confinement.  Here the light-hole (LH) admixture is dominant in
  the lowest spin-split subband, but the heavy-hole (HH) admixture
becomes also important in higher subbands due to the strong HH-LH
  mixing. Then we investigate the time evolution of holes
by numerically solving the fully microscopic
KSBEs in the obtained subbands, with all the scattering,
particularly the Coulomb scattering,
explicitly included. We find that the QWR size influences the
SRT effectively because the SOC and the subband
structure in QWRs depend strongly on the confinement. When the
QWR size increases, the lowest spin-split subband and the
second-lowest spin-split subband will get very close to each other at
an anticrossing point. If the anticrossing is close to the Fermi surface,
the contribution from spin-flip scattering reaches a maximum and,
correspondingly
the SRT will reach a minimum. Moreover, we show that the dependence of
the SRT on confinement size in QWRs behaves oppositely to the trend found in
quantum wells. It is also found that,
 when the QWR  size is very small, the
SRT can  either increase or decrease with hole density, depending on
the spin mixing of the subbands. However, the behavior of holes in QWRs
where the SRT increases or decreases with hole density is
quite different from the one of LHs in quantum wells with small
well width.\cite{Clv}
These features originate from the subband structure of the QWRs and
the spin mixing which give rise to the spin-flip scattering.
The spin mixing and inter-subband
scattering  are  modulated more dramatically in QWRs by changing the
hole distribution in different subbands.
We also investigate the effects of temperature and
initial spin polarization, showing that the inter-subband scattering and
the Coulomb Hartree-Fock contribution can make
a marked contribution to the spin relaxation.

This paper is organized as follows: In Sec.\ II we set up our model and the
KSBEs.  Our numerical results are presented in Sec.\ III. We conclude
in  Sec.\ IV.

\section{Model and KSBE}
Our investigation considers a rectangular $p$-doped (001) GaAs
QWR confined in both $x$ and $y$ directions as shown
schematically in Fig.\ 1. The potential
height of the barrier layer is assumed to be infinite, and the QWR
size in the $x$ ($y$) direction is $a_x$ ($a_y$). Here the $x$,
$y$ and $z$ directions correspond to the [100], [010] and [001]
crystallographic directions, respectively. We assume that the conduction
and valence bands are decoupled, and the effect of the split-off band
is neglected because the spin-orbit split-off energy in bulk GaAs is much
larger than the energy gap between the subbands caused by the confinement
considered here. Then, based on the four-band Luttinger-Kohn model,~\cite{lutt}
the explicit matrix form of the 4$\times$4 bulk-hole Hamiltonian in the basis of
spin-3/2 projection ($J_z$) eigenstates with quantum numbers $+\frac{3}{2}$,
$+\frac{1}{2}$, $-\frac{1}{2}$ and $-\frac{3}{2}$ can be written as\cite{Winkler_book}
\begin{equation}
  \label{Luttinger_Hamiltonian}
  H_{h} = \left( \begin{array}{cccc} H_{hh} & S & R &
      0 \\ S^\dagger & H_{lh} & 0 & R \\ R^\dagger & 0 &
      H_{lh}& -S \\ 0 & R^\dagger & -S^\dagger & H_{hh}  \end{array}
  \right)+  H_{8v8v}^r +   H_{8v8v}^b + V_c({\bf r}),
\end{equation}
where $V_c({\bf r})$ is the hard-wall confinement potential in $x$ and
$y$ directions and
%\begin{subequations}
\begin{eqnarray}
  \label{hh}
  &&  H_{hh} = \frac{1}{2m_0} [(\gamma_1 + \gamma_2)[P_x^2
  + P_y^2] +(\gamma_1 -2 \gamma_2) P_z^2,\\
  \label{lh}
  &&  H_{lh} = \frac{1}{2m_0} [(\gamma_1 - \gamma_2)[P_x^2
  + P_y^2] +(\gamma_1 + 2 \gamma_2) P_z^2,\\
  \label{S}
  &&  S = -  \frac{\sqrt{3} \gamma_3}{m_0} P_z[P_x -  iP_y],\\
  \label{R}
  &&  R = - \frac{\sqrt{3}}{2m_0} \{\gamma_2 [P_x^2 - P_y^2]-
  2 i \gamma_3  P_xP_y \},\\
  \label{Rashba}
  &&  H_{8v8v}^r = \frac{\gamma^{8v8v}_{41}}{\hbar} [J_x (P_y {\cal E}_z -
  P_z {\cal E}_y) + J_y (P_z {\cal E}_x - P_x{\cal E}_z)\nonumber \\
  && \ \ \ \ \ \ \ \ + J_z (P_x {\cal E}_y - P_y {\cal E}_x)], \\
  \label{Dresselhaus}
  &&  H_{8v8v}^b = \frac{b^{8v8v}_{41}}{\hbar^3} \{ J_x [P_x(P_y^2 -
  P_z^2)] + J_y [P_y(P_z^2 -
  P_x^2)] \nonumber \\ && \ \ \ \ \ \ \ \ + J_z [P_z(P_x^2 -
  P_y^2)] \}.
\end{eqnarray}
% \end{subequations}
In these equations, $m_0$  denotes the free electron mass,
$\gamma_1$, $\gamma_2$ and $\gamma_3$ are the Luttinger
parameters, ${\cal E}$ is the  electric field, and $J_i$ are
  spin-3/2 angular momentum matrices.\cite{lutt}  $H_{8v8v}^r$ is the
  SOC arising from the
 structure inversion asymmetry (SIA) and $H_{8v8v}^b$ is the SOC
from the bulk inversion asymmetry (BIA). These two terms turn out to be one or
two orders of magnitude smaller than the intrinsic SOC from the
four-band Luttinger-Kohn Hamiltonian [the first term in Eq.\
(\ref{Luttinger_Hamiltonian})]. This can be seen from
Appendix~A where we present a comparison between spin splittings due
to the SIA and BIA and the splitting from the intrinsic SOC.
Moreover, from the first term in Eq.\ (\ref{Luttinger_Hamiltonian}), one can
 see that the LH spin-up
 states can be directly mixed with the HH states by $S$ and
  $R$, but the mixing between LH spin-up states and LH spin-down
 states has to be mediated by the HH states.
All the mixing is related to the confinement. When the
 confinement decreases, the mixing increases due to the
 decrease of the energy gap between the LH and HH states.

\begin{figure}[t]
  \begin{center}
    \includegraphics[width=3.5cm,height=2.0cm]{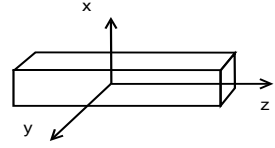}
  \end{center}
  \caption{Schematic geometry of the QWR.}
\end{figure}

We construct the KSBEs by using the nonequilibrium Green function
method as follows:\cite{review}
\begin{equation}
  \label{Bloch_eq}
  \dot{\bf { \rho}}_{{ k}} + \dot{\bf { \rho}}_{{
      k}}|_{coh} +   \dot{\bf { \rho}}_{{ k}}|_{scatt} = 0.
\end{equation}
Here ${\bf { \rho}}_{{ k}}$ represents a single-particle density
matrix of holes with wave vector $k$ along the
  $z$-axis. One can project $\rho_{{ k}}$ in the collinear spin space which is
constructed by basis $\{s\}$, with $\{s\}$  obtained from the eigenfunctions of
the diagonal part of $H_{h}(k)$.
$|s\rangle = |m,n \rangle |\sigma \rangle $ with $ \langle r
|m,n \rangle = \frac{2}{\sqrt{a_x a_y}} \sin(\frac{m\pi
  y}{a_y}) \sin(\frac{n\pi x}{a_x}) e^{i {k}z}$ and $|\sigma
\rangle$ standing for the eigenstates of 
 $J_z$. Then the matrix elements in the
collinear spin space $\rho^c_{k,s_1,s_2}$ is
written as $\rho^c_{k,s_1,s_2} = \langle s_1 |
\rho_k | s_2 \rangle $. Here the superscript ``$c$''
denotes the quantum number distinguishing states in the collinear spin
space. One can also project $\rho_k$ in the ``helix'' spin space which is
constructed by basis $\{\eta\}$ with $\eta$ being the eigenfunctions of
$H_{h}(k)$:
\begin{equation}
\label{spec}
H_{h}(k)|\eta\rangle=E_{\eta,k}|\eta\rangle.
\end{equation}
This basis function is a mixture of
LH and HH states and is $k$ dependent.
Then the matrix elements in the helix spin
space $\rho^h_{{k},\eta,\eta^{\prime}}$ can be written as $\rho^h_{{
    k},\eta,\eta^{\prime}} = \langle \eta |\rho_k|
\eta^{\prime} \rangle$, with the superscript ``$h$''
denoting  the helix spin
space. The density matrix in the helix spin space can be
transformed from that in the collinear one by a unitary
transformation: $ \rho^h_k = U^{\dagger}_k\rho_{k}^c U_k $,
where $U_k(i,\alpha) = \eta^i_{\alpha}(k)$ with
$\eta^i_{\alpha}(k)$ being the $i$th element of the $\alpha$th eigenvector
after the diagonalization of $H_h(k)$.

In this paper we project the density matrix in the helix spin space\cite{Cheng2}
and then the coherent terms can be written as
\begin{eqnarray}
  \label{coh}
\dot{\rho}^h_{{k}}|_{coh} &=& - i \Big[ \sum_{\bf Q} V_{\bf Q}
U^{\dagger}_{k}I_{\bf Q} U_{k-q} { \rho}^h_{k-q}
U^{\dagger}_{k-q}I_{\bf -Q} U_{k},  \rho^h_k
\Big] \nonumber \\
&&\mbox{}-i \Big[ U^{\dagger}_{k}H_{h}(k)U_{k} ,  \rho^h_k
\Big],
\end{eqnarray}
where $[A,B] = AB - BA$ denotes the commutator, and
$I_{\bf Q} $  is the form factor in the collinear spin space with wave
vector ${\bf Q} \equiv (q_x, q_y, q)$.
The first term in Eq.\ (\ref{coh}) is the
Coulomb Hartree-Fock term, and the second term is the contribution
from the intrinsic SOC from the  Luttinger-Kohn Hamiltonian.
$I_{\bf Q} $ can be written as  $I_{{\bf Q},s_1,s_2} = \langle s_1 | e^{i
  {\bf Q} \cdot {\bf r}} | s_2 \rangle  = \delta_{\sigma_1,\sigma_2}
F(m_1,m_2,q_y,a_y)F(n_1,n_2,q_x,a_x)$, with
\begin{widetext}
\begin{equation}
F(m_1 , m_2,q,a)= 2 i a q [e^{i a q} \cos{\pi (m_1 -
 m_2)} -1 ] 
 \left[\frac{1}{\pi^2(m_1 - m_2)^2 -
 a^2 q^2}% \nonumber \\ &&
- \frac{1}{\pi^2(m_1 + m_2)^2 - a^2 q^2}\right].
\end{equation}
For small spin polarization, the contribution from the
Hartree-Fock term in the coherent term is negligible\cite{Weng,schu}
 and the spin precession  is determined
by the SOC,
$\dot{\rho}^h_{{k},\eta,\eta^{\prime}}|_{coh} = -i\rho^h_{{k},\eta,\eta^{\prime}}(E_{\eta,k} -
  E_{\eta^{\prime},k})$, which is proportional to the energy gap between
  $\eta$ and $\eta^{\prime}$ subbands.

The scattering terms include the
hole-nonmagnetic-impurity,  hole-phonon
and hole-hole Coulomb scatterings. In the helix spin space,
The scattering terms are given by
%\begin{widetext}
\begin{eqnarray}
  \dot{\rho}^h_{{k}}|_{scat} &=& \pi N_i \sum_{{\bf
      Q},\eta_1,\eta_2} |U^i_{\bf Q}|^2
  \delta(E_{\eta_1,k-q} - E_{\eta_2,k})  U^{\dagger}_{k}I_{\bf Q} U_{k-q}
  [(1-{ \rho}^h_{k-q})T_{k-q,\eta_1}
  U^{\dagger}_{k-q}I_{-{\bf Q}} U_{k}T_{k,\eta_2} {
    \rho}^h_k \nonumber \\
&&\mbox{}-
  \rho^h_{k-q} T_{k-q,\eta_1} U^{\dagger}_{{k-q}}I_{-{\bf Q}} U_{k}T_{k,\eta_2} (1-\rho^h_k)]
  + \pi \sum_{{\bf Q},\eta_1,\eta_2,\lambda} |M_{{\bf Q},\lambda}|^2
  U^{\dagger}_{k}I_{\bf Q} U_{k-q} \{
  \delta(E_{\eta_1,k-q} - E_{\eta_2,k} +
  \omega_{{\bf Q},\lambda} )\nonumber \\
&&\mbox{}\times [(N_{{\bf Q},\lambda} +1)(1-{
    \rho}^h_{k-q})  T_{k-{q},\eta_1}U^{\dagger}_{k-q}I_{-{\bf Q}} U_{k}T_{{
      k},\eta_2} { \rho}^h_k - N_{{\bf Q},\lambda} \rho^h_{k-q}  T_{{
      k}-q,\eta_1} U^{\dagger}_{{\bf
      k-q}}I_{-{\bf Q}} U_{k}T_{k,\eta_2}(1-\rho^h_k)] \nonumber \\
&&\mbox{} +\delta(E_{\eta_1,k-q} - E_{\eta_2,k} -
  \omega_{{\bf Q},\lambda} ) [N_{{\bf Q},\lambda}(1-{
    \rho}^h_{k-q})  T_{{k}-q,\eta_1}U^{\dagger}_{k-q}I_{-{\bf Q}}
U_{{k}}T_{k,\eta_2}{ \rho}^h_k \nonumber \\ && - (N_{{\bf
 Q},\lambda}+1) \rho^h_{k-q}  T_{{
      k}-q,\eta_1}U^{\dagger}_{k-q}I_{-{\bf Q}} U_{k}T_{k,\eta_2}(1-\rho^h_k)]
  \}  \nonumber \\
&&\mbox{}+ \pi \sum_{{\bf Q},k^{\prime}}
  \sum_{\eta_1,\eta_2,\eta_3,\eta_4} V_{\bf Q}^2 U^{\dagger}_{k}I_{\bf Q} U_{k-q}
 \delta(E_{\eta_1,{
      k}-q} - E_{\eta_2,k} + E_{\eta_3,k^{\prime}} -
  E_{\eta_4, k^{\prime} -q})
%  U^{\dagger}_{k}I_{\bf Q} U_{k-q}
 \nonumber \\
&&\mbox{}\times \{
  (1-{\rho}^h_{k-q})T_{k-q,\eta_1}
  U^{\dagger}_{k-q}I_{-{\bf Q}} U_{k} T_{{
      k},\eta_2}{   \rho}^h_k
  \mbox{Tr}[(1-\rho^h_{k^{\prime}})T_{\eta_3,k^{\prime}}
  U^{\dagger}_{k}I_{\bf Q} U_{k-q}T_{k^{\prime} -q,\eta_4}\rho^h_{k^{\prime}
    -q}
  U^{\dagger}_{k-q}I_{-{\bf
Q}} U_{k}]  \nonumber \\
&&\mbox{}- \rho^h_{{\bf
      k}-q} T_{k-q,\eta_1} U^{\dagger}_{k-q}I_{-{\bf Q}} U_{k} T_{{
      k},\eta_2}(1-\rho^h_k)
  \mbox{Tr}[\rho^h_{k^{\prime}}T_{\eta_3,k^{\prime}}
  U^{\dagger}_{k}I_{\bf Q} U_{k-q} T_{k^{\prime} -q,\eta_4} (1-\rho^h_{
      k^{\prime}-q})
  U^{\dagger}_{k-q}I_{-{\bf Q}} U_{k}]\} \nonumber \\
&&\mbox{}  + h.c.
\label{scat}
\end{eqnarray}
\end{widetext}
in which $T_{k,\eta}(i,j) =  \delta_{\eta i} \delta_{\eta j}
$. $V_{\bf Q}$ in Eq.\ (\ref{scat}) reads $ V_{\bf Q} =
4 \pi e^2 /[\kappa_0 (q^2 + q_{\|}^2 + \kappa^2)]$, with
$\kappa_0$ representing the static dielectric constant and $\kappa^2 =
4\pi e^2 N_h /(k_B T \kappa_0 a^2)$ standing for the Debye screening
constant. $N_i$ in  Eq.\ (\ref{scat}) is the impurity density and  $
|U^{i}_{\bf Q}|^2 = \{ 4\pi Z_i  e^2 /[\kappa_0 (q^2 + q_{\|}^2 +
\kappa^2)]\}^2$ is the impurity potential with $Z_i$ standing for the
charge number of the impurity.  $|M_{{\bf Q},\lambda}|^2$ and $N_{{\bf
    Q},\lambda} = [\mbox{exp}(\omega_{{\bf Q},\lambda} /k_B T) - 1]^{-1}$ are
the matrix element of the hole-phonon interaction and the Bose
distribution function  with phonon energy spectrum $\omega_{{\bf
    Q},\lambda}$ at phonon mode $\lambda$ and wave vector ${\bf Q}$,
respectively. Here the hole-phonon scattering includes the
hole--LO-phonon  and hole--AC-phonon scatterings with the
explicit expressions of $|M_{{\bf Q},\lambda}|^2$ can be found in
Refs.\ \onlinecite{Weng, Zhou}.

It is noted that in the scattering terms Eq.\ (\ref{scat}),  the energy spectra
$E_{\eta,k}$ are from Eq.\ (\ref{spec}) with full SOC included.
As discussed by Cheng and Wu, this spectrum leads to the
so called  helix statistics in the equilibrium,\cite{Cheng2}
i.e., the Fermi distribution with SOC included in the energy spectrum.
In most of our previous
works,\cite{Wu1,Wu2,Wu3,Weng,Wu5,Cheng,Clv,Zhou} the energy spectra in the
scattering terms do not include the SOC, i.e., the energy spectra
from the Hamiltonian without the SOC. The corresponding equilibrium statistics is
referred to as the collinear statistics.\cite{Cheng2}
It has been demonstrated that when the SOC is weak,
the collinear statistics is good
enough. However, the SOC for holes in QWR system can be very strong.
Therefore, it is important to adopt  the
helix statistics in this investigation.

Finally we comment on the  reason we solve the KSBEs in the helix spin space
$\rho^h_k$. This is because the numerical calculation
becomes faster in the helix spin space. This can be understood from the fact  that
even the lowest helix subband is a mixture of many collinear basis states $\{s\}$
due to the strong SOC.
Therefore, a large number of basis states have to be included in the calculation
if we project the KSBE in the collinear spin space.

\section{Numerical Results}

We first solve Eq.\ (\ref{spec}) to obtain the subband
structure. In Fig.\ 2 we show six typical energy spectra
for different confinements. Each subband is denoted as
  $l+$ ($l-$) if the dominant spin component  is the
  spin-up (-down) state.
One can see from Fig.~2 that $1+$ and $1-$ are very close
  to each other, so are the subbands $2\pm$. The spin-splitting
  between them is mainly caused by the SOC arising from BIA, for that the
  spin-splitting caused by the SOC arising from SIA is three orders of
  magnitude smaller
 than the diagonal terms in Eq.\ (\ref{Luttinger_Hamiltonian}) and
  can not be seen in Fig.~2. The spin-splitting caused by the BIA is
  proportional to $(P_x^2 - P_y^2)$, which disappears when the
confinement in $x$ and $y$ directions are symmetrical. Therefore,
 $l\pm$ are almost degenerate when $a_x = a_y =
  10$~nm. If one excludes the SOC from the BIA and SIA, $l\pm$
are always degenerate because of the Kramers degeneracy.
One also observes that when $a_x$ gets larger, the subbands are closer to
each other. Especially, in the case of
$a_x=a_y=10$~nm, there are anticrossing points due to the
HH-LH mixing in the Luttinger Hamiltonian.
When $a_x$ keeps on increasing,
the anticrossing
point at small $k$ between the $1\pm$ and $2\pm$ gradually
 disappears. However, at large $k$ region,
the lowest two subbands become very close to each other. These will lead
to significant effect on SRT.
\begin{figure}[thb]
  \begin{center}
    \includegraphics[width=8.5cm,height=3.7cm]{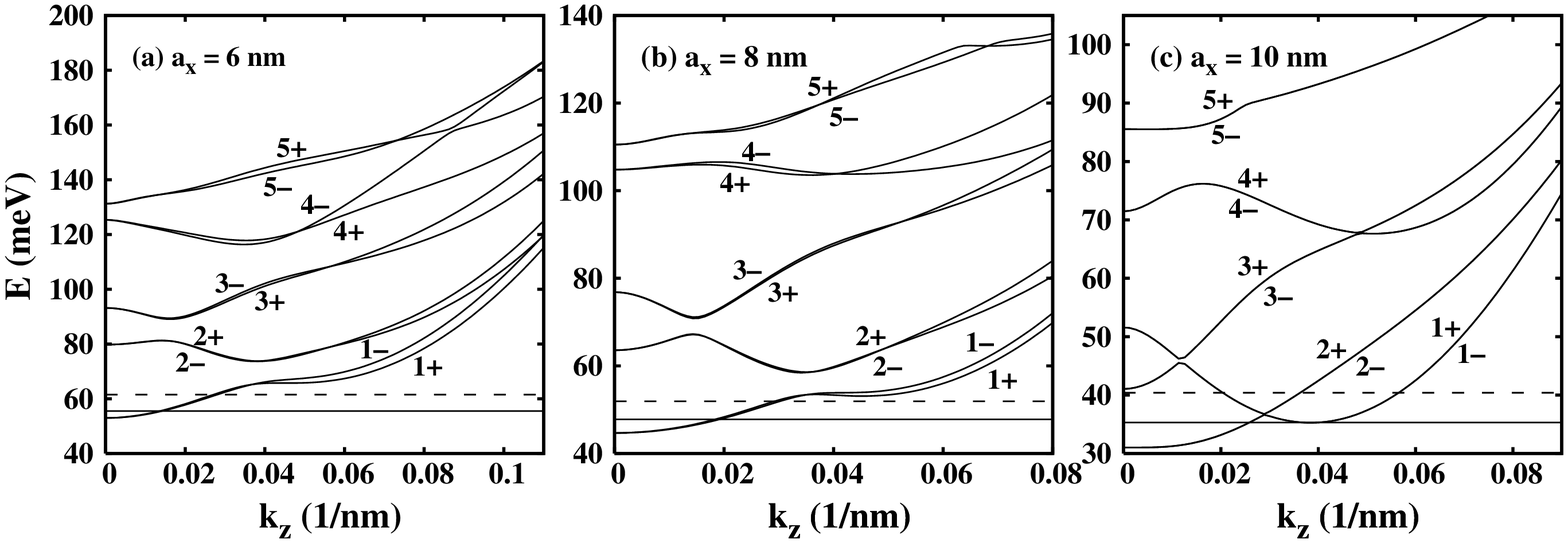}
  \end{center}
  \begin{center}
    \includegraphics[width=8.5cm,height=3.7cm]{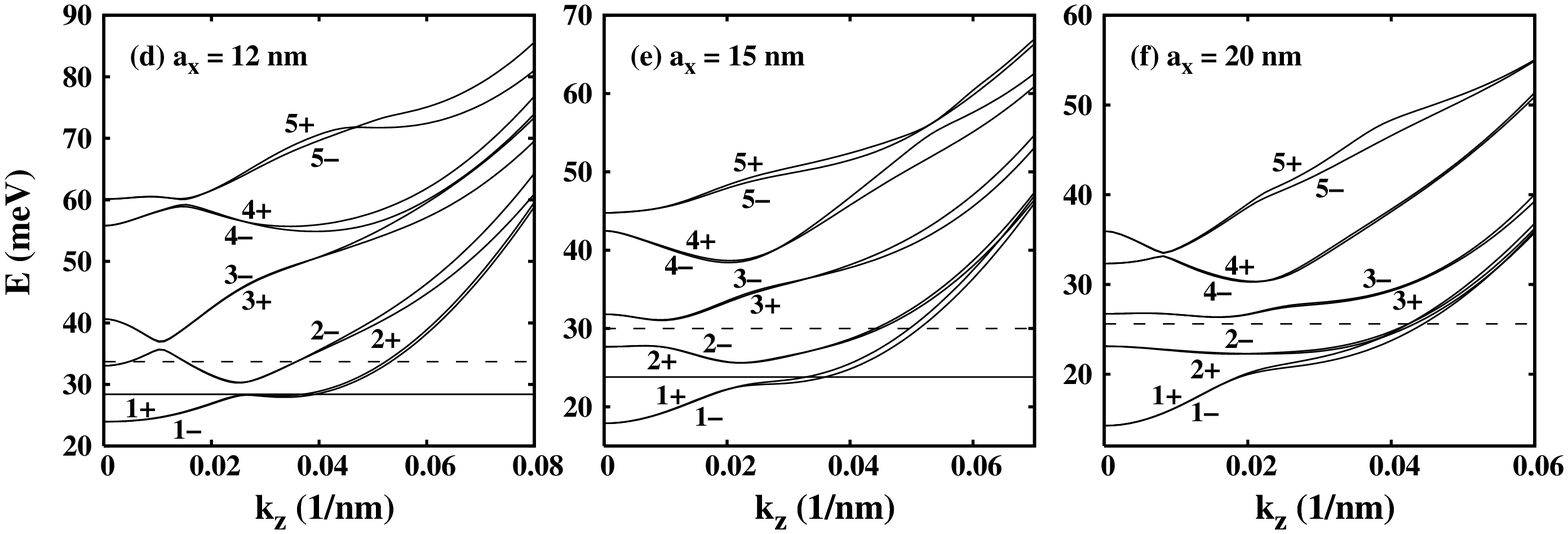}
  \end{center}
  \caption{Typical energy spectra for (a) $a_x=6$~nm; (b) $a_x=8$~nm; (c)
    $a_x=10$~nm; (d) $a_x=12$~nm; (e)  $a_x=15$~nm; and (f)
    $a_x=20$~nm. $a_y=10$~nm.
${\langle E\rangle}$ at $T = 20$~K is also
      plotted: solid line for $N_h = 4 \times 10^5 $~cm$^{-1}$ and
    dashed line for $N_h = 2 \times 10^6 $~cm$^{-1}$. }
\end{figure}

In order to show the situation of hole's population in these
  subbands clearly, We introduce a quantity $\langle E \rangle$, with
\begin{equation}
{\langle E\rangle} =\frac{\sum_{l}\int^{+\infty}_{-\infty} dk
(\rho^h_{k,l+,l+}-\rho^h_{k,l-,l-}) (E_{l+,k}+E_{l-,k})}{2\sum_{l}\int^{+\infty}_{-\infty} dk
  (\rho^h_{k,l+,l+}-\rho^h_{k,l-,l-})},
\end{equation}
to represent the energy region where spin precession and relaxation
between the $+$ and $-$ bands mainly take place.
Due to the small spin polarization, $\langle E \rangle$ is approximately equal
to the Fermi energy at very low temperature. In Fig.~2 we plot
$\langle E\rangle$ for $N_h = 4 \times 10^5$~cm$^{-1}$ and $2 \times
10^6$~cm$^{-1}$ at $T=20$~K. It is seen that $\langle E \rangle$ only
intersects with the $1\pm$ and $2\pm$ subbands, which means holes are
mainly populated in $1\pm$ and $2\pm$ subbands.\cite{comment}
Therefore,
only the $1\pm$ and $2\pm$ subbands are taken into account in the present
investigation. Higher subbands should be included  if
one considers  higher hole density, higher
temperature, or larger QWR size. It is further noticed
 that the dominant spin component in $1+$ ($1-$) state is the spin-up
(spin-down) LH state. At small $k$, the spin-up
 (spin-down) LH admixture remains at more than 90\ \%. Moreover,
the HH-LH mixing in $2\pm$ subbands is much stronger.

After the energy spectrum is obtained, we numerically solve the KSBEs
and obtain the
temporal evolution of the hole density matrix $\rho^h_{k}(t)$ in helix
spin space.
Then we project $\rho^h_{k}(t)$ back into the collinear spin
space $\rho^c_{k}(t)$ and obtain the temporal evolution of the spin polarization
\begin{equation}
{\bf S}^c_k(t) = \mbox{Tr}[\rho_k^c(t) {\bf J}],
\end{equation}
 in which ${\bf J}$
is the operator for spin-3/2 angular momentum, written as a matrix in the basis
of $z$-projection eigenstates with eigenvalues $ m = +3/2, +1/2, -1/2, -3/2$.
We include the hole-phonon and  hole-hole scatterings throughout our
computation. The material parameters of GaAs in our calculation are the
same as those used in Refs.~\onlinecite{Clv,Zhou}.
The initial condition at $t=0$ is set to be spin
polarized with a small initial spin polarization ${P}$
which is given, in the helix spin
space, by $P=(N_{1+}+N_{2+}-N_{1-}-N_{2-})/N_h$ where $N_h$ is the
total hole density. Therefore, we have initial spin polarization
along all directions
in the collinear spin space. Then as discussed in the previous
papers,\cite{Weng}
the  SRT $\tau$ can be defined by the slope
of the envelope of the  spin polarization along the $z$-axis:
\begin{eqnarray}
  \label{sum_S}
S^c_z = \sum_k {S}^c_{k,z}(t).
\end{eqnarray}

\subsection{Spin relaxation mechanisms}

There are three mechanisms leading to spin relaxation. First, the
spin-flip scattering, which includes the scattering between $l+$ and
$l-$ subbands and the scattering between $l+$ and $l^{\prime}-$
subbands ($l\ne l^{\prime}$), can cause spin relaxation. The SRT
decreases with the spin-flip scattering, with the scattering strength
being proportional to the spin mixing of the helix subbands. Second,
because of the coherent term $\dot{\rho}^h_{{k}}|_{coh}$, there is a
spin precession between different subbands. The frequency of this spin
precession depends on $k$ and this dependence serves as inhomogeneous
broadening. As shown in 
Refs.~\onlinecite{review,Wu1,Wu2,Wu3,Weng,Wu5}, in the presence of the
inhomogeneous broadening, even the spin-conserving scattering can
cause irreversible spin relaxation.  As a result, the spin-conserving
scattering, i.e., the scattering between $l+$ and $l^{\prime}+$ and
the scattering between $l-$ and $l^{\prime}-$, can cause spin
relaxation along with the inhomogeneous broadening. At last, the
spin-flip scattering along with the inhomogeneous broadening can also cause an
additional spin relaxation.  

It is seen from Fig.\ 2(a) that when $N_h = 4 \times 10^5 $~cm$^{-1}$
and $a_x = 6$~nm,
$\langle E \rangle$ only intersects  with the $1\pm$ subbands and is
far away from the $2\pm$ subbands. Therefore,
holes populate the $1\pm$  subbands only.
 As pointed out before, the coherent term $\dot{\rho}^h_{k,1+,1-}|_{coh}$ is
proportional to $(E_{1+,k}-E_{1-,k})$. As holes are only
populating states in the small $k$ region where the
spin splitting between $1\pm$ is negligible,  the spin
precession between these two states, and thus the inhomogeneous
broadening, is very small. Consequently  the main spin-relaxation mechanism
is due to the spin-flip scattering, i.e., the scattering between
$1\pm$ subbands. 

In the case of larger $a_x$ and $N_h$ as shown in Fig.~2(c)-(f),
where $\langle E \rangle$ is close to or intersects with the $2\pm$ subbands,
holes populate both the $1\pm$ and $2\pm$ subbands. The spin-flip
scattering here includes the scattering between 
$1\pm$ states, the scattering between $2\pm$ states and the spin-flip
scattering between $1\pm$ and $2\pm$ subbands. 
This spin-flip scattering is still found to be the main spin
relaxation mechanism. Besides, differing from the case of
Fig.\ 2(a), the coherent term $\dot{\rho}^h_{k,1\pm,2\pm}|_{coh}$ 
is proportional to the energy gap between $1\pm$ and $2\pm$, and it is
much larger than $\dot{\rho}^h_{k,1+,1-}|_{coh}$. As a result, there
is a much stronger spin precession between $1\pm$ and $2\pm$ subbands
with a frequency depending on $k$, and the inhomogeneous broadening
caused by this precession along
with both the spin-conserving scattering and the spin-flip scattering
can make a considerable contribution to the spin relaxation.

\subsection{Wire width dependence of the SRT}

\begin{figure}[th]
  \begin{center}
    \includegraphics[width=9cm]{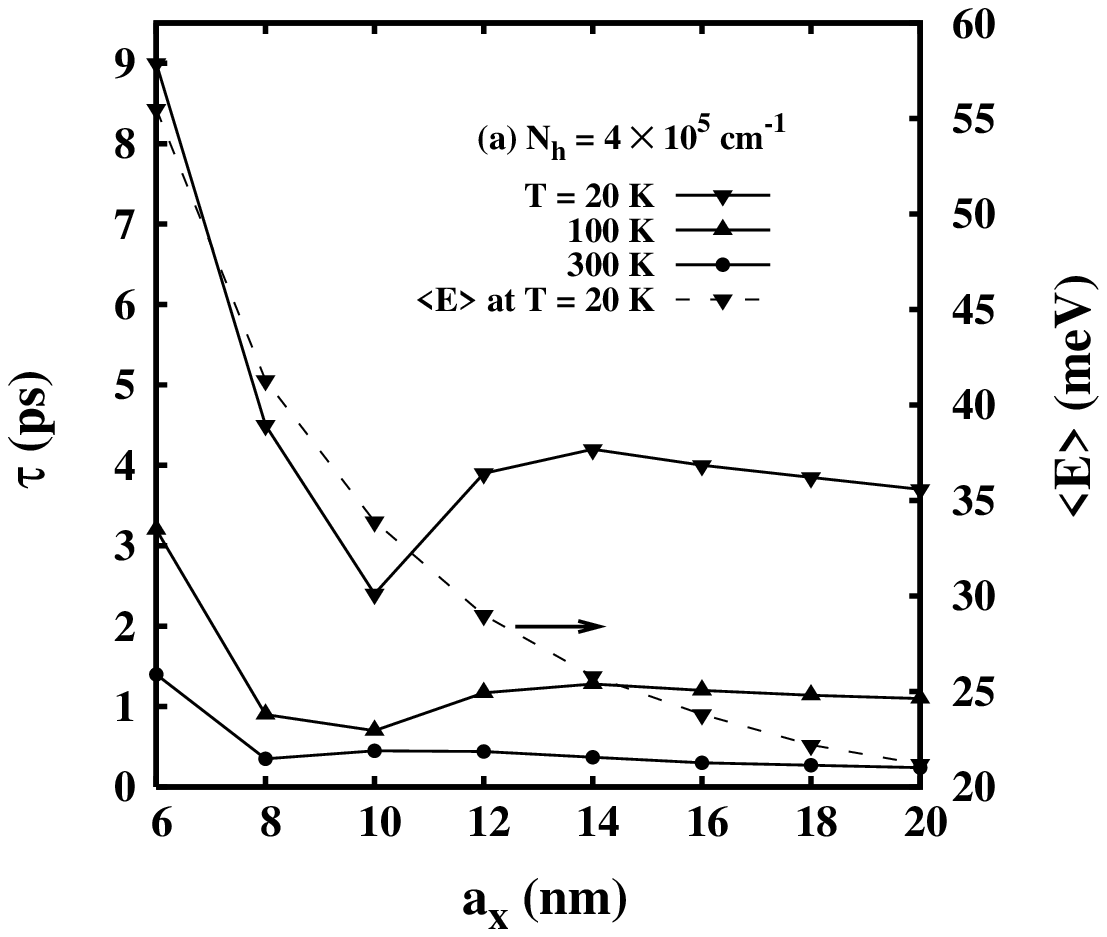}
  \end{center}
  \begin{center}
    \includegraphics[width=9cm]{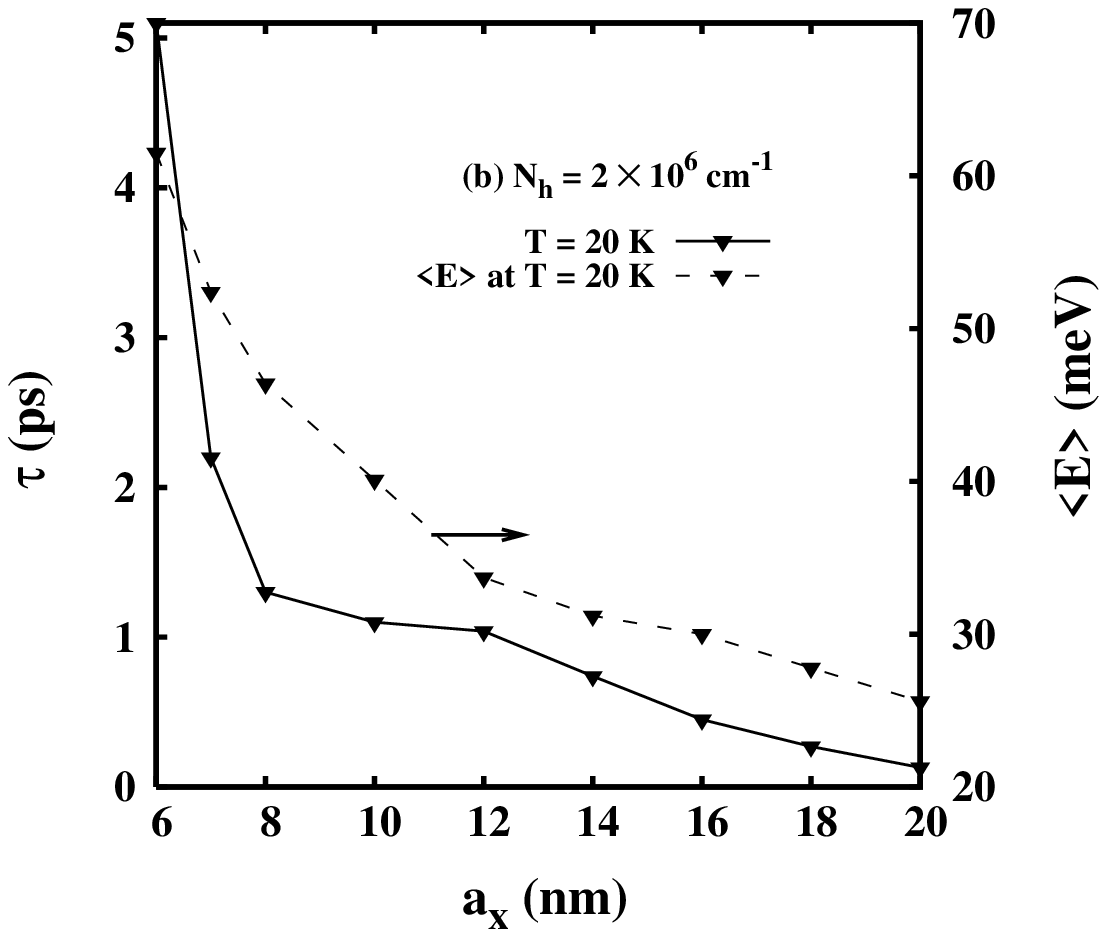}
  \end{center}
  \caption{SRT $\tau$  {\em vs.} the QWR width in $x$
    direction $a_x$ for (a) $N_h = 4 \times
10^5$~cm$^{-1}$ at different temperatures and (b) $N_h = 2 \times
10^6$~cm$^{-1}$ at $T = 20$~K. $a_y=10$~nm.}
\end{figure}

In Fig.\ 3 we plot the SRT as a function of the QWR width in $x$
direction, $a_x$, for various temperatures. Here $a_y = 10$\ nm,
the hole density is taken to be $N_h = 4 \times 10^5$~cm$^{-1}$
in Fig.\ 3(a) and $N_h = 2 \times 10^6$~cm$^{-1}$
in Fig.\ 3(b). It is seen from Fig.\ 3(a) that in the case of low
density and $T = 20$\ K, the SRT first decreases with $a_x$ when $a_x <
10$\ nm, then increases with $a_x$ when 10~nm $<a_x<14$\ nm, and
 finally decreases with $a_x$  when $a_x > 14$\ nm. To understand this
behavior, let's look at the energy spectra
for a wire with $a_x=6$\ nm and $10$\ nm
in Fig.\ 2. When the wire width increases, the energy gap between the
$1\pm$ and $2\pm$ becomes smaller, and the spin mixing
in the helix subbands increases.
Therefore, the contribution from all of the spin-flip
scattering increases, and the SRT decreases with increasing $a_x$.
From this point of view, one can expect a minimum of SRT when
all of the spin mixing reaches a maximum at the Fermi surface
as we can approximately make the assumption that the spin
relaxation occurs mainly around the Fermi surface.
In order to show this effect, let's look at $\langle E \rangle$ at $T
= 20$~K for $N_h = 4 \times 10^5 $~cm$^{-1}$ in Fig.\ 2(c). One can
see that in the case of $a_x = a_y = 10$~nm, the
lowest two subbands have an anticrossing and $\langle E \rangle$
is very close to the anticrossing point. From our calculation, we find
that all of the spin mixing, including the HH-LH mixing, the mixing
between the LH up states and the LH down states, and the mixing
between the HH up states and the HH down states, reaches a maximum
because of the anticrossing. As a result, this
anticrossing point leads to a strong spin-flip scattering and
accounts for the minimum of
SRT in Fig.\ 3(a) at $T= 20$\ K as expected. When $a_x$ keeps on increasing,
 $\langle E \rangle$ will move into the larger-$k$ region and the
anticrossing point will disappear gradually as shown in Fig.\ 2, and
the energy gap
between the lowest two helix subbands at
the Fermi surface gets larger. Therefore, the spin mixing
at the Fermi surface becomes smaller, and the SRT
will slightly increase with $a_x$. However, when $a_x\ge 14$\ nm, the
effect of reducing the 
energy gap between the lowest two subbands and increasing the spin mixing
are more important and the SRT decreases with $a_x$ again.

We also plot the SRT as a function of $a_x$ at $100$\ K and $300$\ K in
Fig.\ 3(a). One finds that the SRT decreases with temperature. This
could be understood as follows: firstly, when the temperature increases, 
holes are populating the higher-$k$ states, for which all of the
spin mixings in the $1\pm$ and $2\pm$ subbands are stronger; secondly, the
strength of total scattering is enhanced. Both effects
increase the contribution of spin-flip scattering and speed up the
spin relaxation.

It is seen from Fig.\ 2(f) that the lowest two subbands are very
close to each other in large $k$ region. Therefore, one can
expect a very short SRT when the Fermi surface enters this
region as the spin mixing here is very large and the
contribution from the spin-flip scattering
could be very strong. In order to show this effect, we take the hole
density to be $N_h = 2 \times 10^6$~cm$^{-1}$ to place the Fermi surface
into the larger $k$ region in Fig.\ 3(b).
It is seen from the figure that the SRT decreases monotonically with
$a_x$. This could be easily understood from the fact that the LH-HH
mixing increases with $a_x$ due to the decrease of the energy gap
between the LH and HH states. In the case of $a_x = 20$~nm, the SRT
is nearly two orders of magnitude smaller than the SRT
at $a_x = 6$~ nm as expected.

\begin{figure}[t]
  \begin{center}
    \includegraphics[width=9cm]{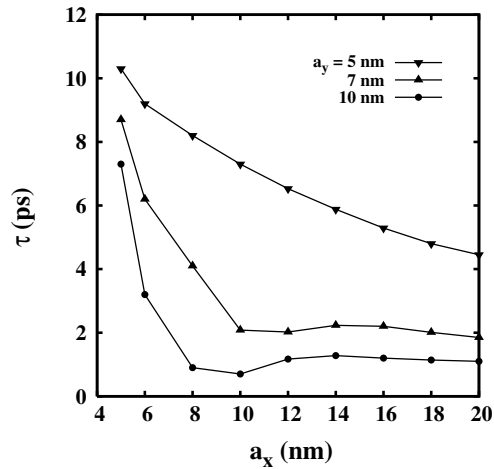}
  \end{center}
  \caption{SRT $\tau$  {\em vs.} the QWR width in $x$
    direction at different $a_y$ at $N_h = 4 \times 10^5$~cm$^{-1}$
    and $T = 100$~K.}
\end{figure}

In Fig.\ 4 we plot the SRT as a function of $a_x$ with different $a_y$
at $N_h = 4 \times 10^5$~cm$^{-1}$ and $T = 100$~K. It
is seen that, when $a_y = 5$\ nm, the SRT decreases
monotonically with $a_x$. This is because when the confinement is strong,
there is no anticrossing point between the lowest two subbands in the
region where holes are distributed. As a result, the
SRT decreases with $a_x$ because of the effect of reducing the
energy gap between the lowest two subbands. When $a_y$ is increased, the
anticrossing point appears and the SRT shows a minimum similar to the
case shown in Fig.\ 3(a).

\subsection{Hole density and temperature dependence of SRT}

\begin{figure}
  \begin{center}
    \hspace{-1.5cm}\includegraphics[width=9cm]{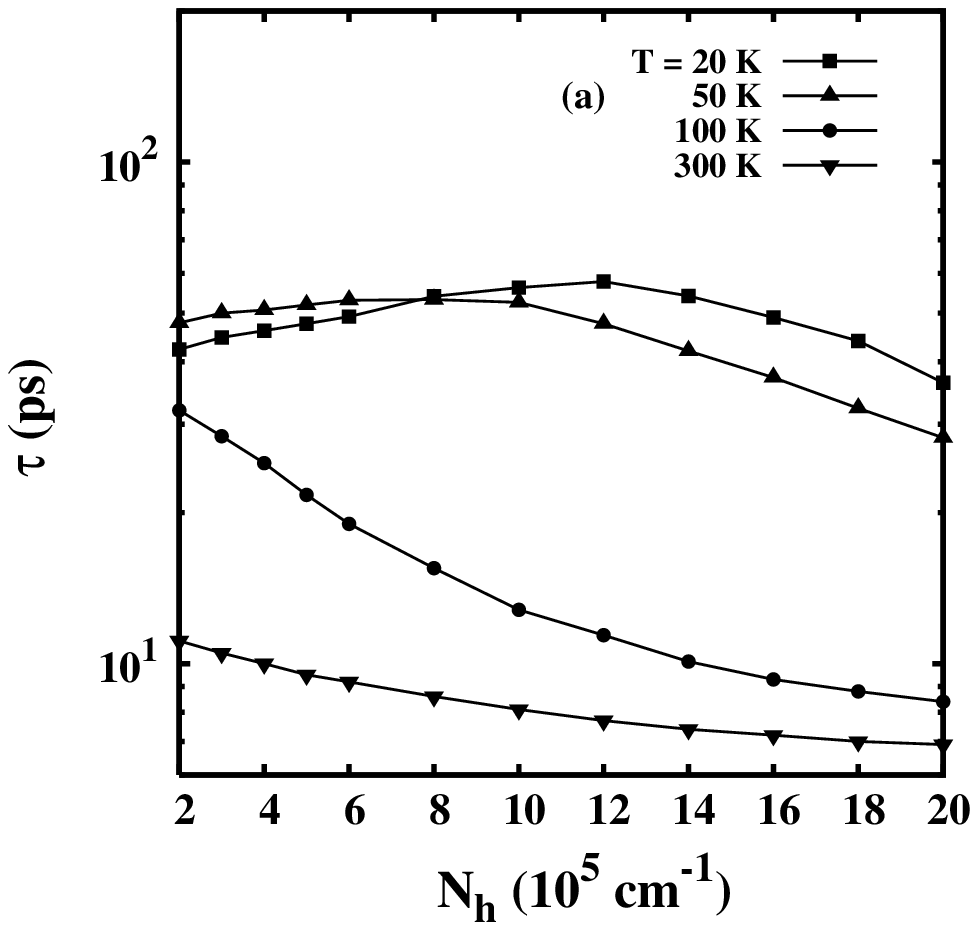}
  \end{center}
  \vskip 0.5pc
  \begin{center}
    \hspace{0.0cm}\includegraphics[width=9cm]{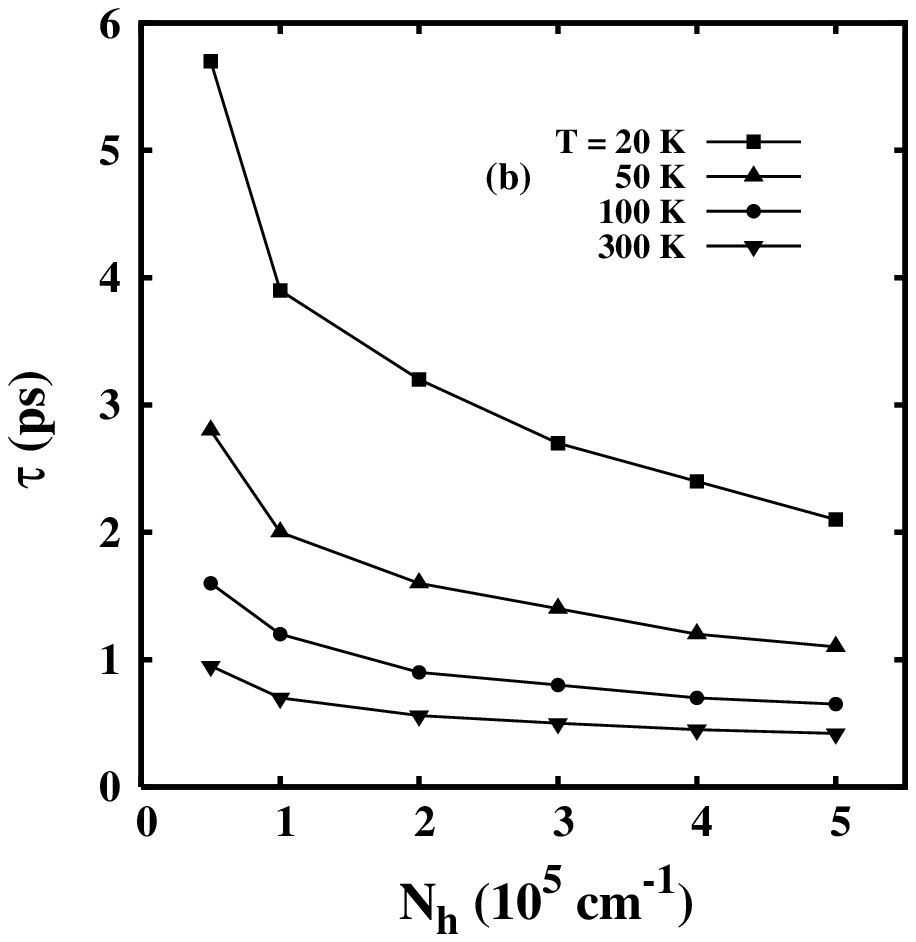}
  \end{center}
  \vskip 0.5pc
  \begin{center}
    \hspace{0.0cm}\includegraphics[width=9cm]{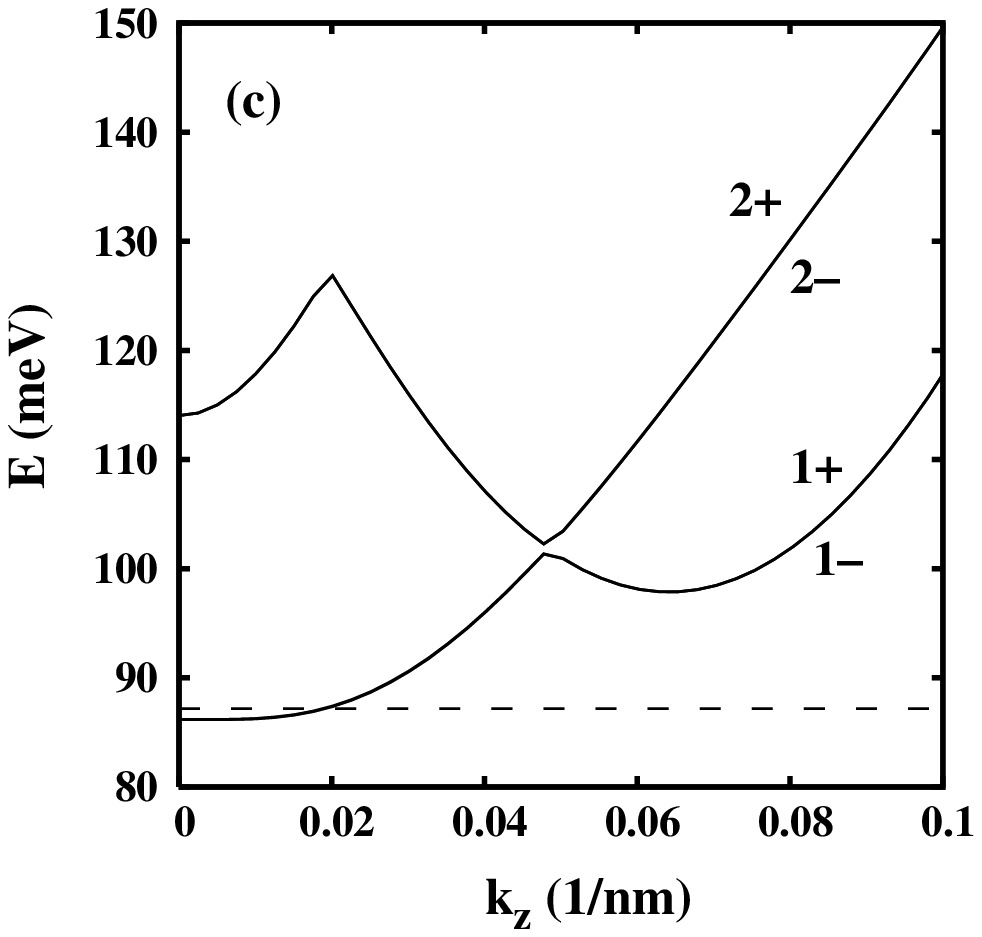}
  \end{center}
  \caption{SRT {\em vs.} the hole density at different temperatures.
    (a) $a_x = a_y = 6$\ nm; (b) $a_x = a_y = 10$\ nm. The energy
    spectrum for $a_x = a_y = 6$\ nm is shown in (c). }
\end{figure}

Now we turn to study the hole-density dependence of the SRT
at different temperatures and confinement sizes.
In Fig.\ 5(a) we plot the SRT as a function of $N_h$ at various
temperatures and $a_x = a_y = 6$\ nm. From the figure one can
see that, when $T \le 50$\ K, the SRT first increases then decreases
with $N_h$. To understand this behavior, let's look at the energy
spectrum for $a_x = a_y = 6$~nm shown in Fig.\ 5(c). The dashed
line in Fig.\ 5(c) represents $\langle E \rangle$ for $N_h = 12 \times
10^5$~cm$^{-1}$ at $T=20$~K. One can see that the energy gap between
$1\pm$ and $2\pm$ is large because of the small QWR size.
When the temperature is low, $\langle E \rangle$ only
intersects  with the $1\pm$ subbands and is far away from the $2\pm$
subbands. Therefore, holes populate the $1\pm$  subbands only. The
main spin relaxation mechanism is from the spin-flip
scattering, i.e., the scattering between $1\pm$ subbands. Furthermore,
it can be seen that the $2\pm$ subbands have a maximum at the wavenumber
where the $1\pm$ subbands intersect with $\langle E \rangle$ for $N_h
= 12 \times 10^5$~cm$^{-1}$ at $T=20$~K.
In the region where $k$ is smaller than the intersection point of $\langle E
\rangle$ and the $1\pm$ subbands, the energy gap between $1\pm$ and $2\pm$
subbands increases, and our calculation shows that the spin mixing of the
$1\pm$ states decreases. Therefore, when $T=20$~K and $N_h \le 12 \times
10^5$~cm$^{-1}$, the spin mixing at the Fermi surface decreases with
increase of $N_h$, and the SRT increases with $N_h$ because of the
decrease of the spin-flip scattering. In the region where $k$ is larger
than the intersection point, the energy gap between $1\pm$ and $2\pm$
subbands decreases, and the spin mixing of the $1\pm$ states
increases. As a result, the SRT increases with $N_h$ when $N_h > 12
\times 10^5$~cm$^{-1}$. The case of $T=50$~K is similar to that of
$T=20$~K, but the holes are distributed in a wider $k$ region and
reach the maximum of the $2\pm$ subbands at a smaller
$N_h$. Therefore, the SRT begins to
decrease at $N_h = 8 \times 10^5$~cm$^{-1}$. One can further see that the
SRT for $T=20$~K is smaller than that for $T=50$~K at small $N_h$
but larger than it at large $N_h$. This can be easily understood 
as when the Fermi wavevector is smaller than the wavevector where the
maximum of the $2\pm$ subbands occurs, the spin mixing decreases with
$T$ and the SRT increases with $T$. Otherwise, the SRT decreases with $T$.

We also plot the SRT as a function of $N_h$ at $100$\ K and $300$\ K in
Fig.\ 5(a). One finds that the SRT decreases with $N_h$. This
can be understood as follows: firstly, holes are populated at high-$k$
states that are larger than 
the wavevector where the maximum of the $2\pm$ subbands occurs, and the
spin mixing increases with $N_h$; secondly, due to the larger
temperature, the holes are also distributed in $2\pm$ states.
Therefore, the spin-flip scattering includes not only the scattering
between $1\pm$ subbands, but also
the inter-subband spin-flip scattering, i.e., the
spin-flip scattering between $1\pm$ and $2\pm$ subbands. The
strength of this inter-subband spin-flip scattering is enhanced with
the increase of $N_h$ because of the increase of the hole population in
$2\pm$ states. Both effects increase the contribution of spin-flip
scattering and boost the spin relaxation.

The results shown in Fig.\ 5(a) are quite different compared with 
those of LHs in quantum
wells with small well width where the SRT decreases monotonically with
$N_h$ at low temperature but increases with $N_h$ at high
temperature.\cite{Clv} The difference originates from
the fact that the
energy spectrum of the QWR is modulated dramatically by the
QWR size, and one can modulate the spin mixing
strength by changing the region where holes are distributed.
In order to better show this modulation, we also plot the results
of $a_x = a_y = 10$\ nm in Fig.\ 5(b). In this situation the energy
gap between the lowest two subbands gets smaller as shown in Fig.\
2(c), and consequently both the spin mixing and the strength of 
the inter-subband spin-flip scattering increase with
$N_h$ like the case of $T \ge 100$~K in Fig.\ 5(a). Therefore, the SRT
decreases with $N_h$ as expected. The case of $N_h > 5 \times
10^5$~cm$^{-1}$ is not calculated in Fig.\ 5(b) as higher subbands should
be included when $N_h > 5 \times 10^5$~cm$^{-1}$ and $T \ge 200$~K,
while only the $1\pm$ and $2\pm$ subbands are taken into account in our
investigation.

\begin{figure}
  \begin{center}
\includegraphics[width=9cm]{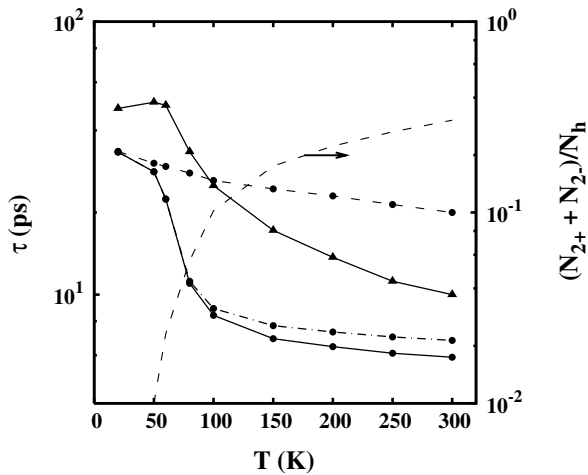}
  \end{center}
  \caption{SRT $\tau$  {\em vs.} temperature $T$.
    {$\blacktriangle$}: $N_h = 4 \times 10^5$ m$^{-1}$; {$\bullet$}:
    $N_h = 2 \times 10^6$ m$^{-1}$. The QWR
    size is $a_x = a_y = 6$\ nm. The solid curve is the result
    from the full calculation; the dashed curve is the result without the
    inter-subband hole-phonon scattering and inter-subband hole-hole
    scattering; the chain curve is the result without the coherent
    term. We also plot the hole distribution of the $2\pm$ subbands.
    Note the scale of the distribution is on the right side of the figure.}
\end{figure}

To see more detail of how the temperature affects the spin relaxation,
we plot in Fig.\ 6 the SRT as a function of $T$ with the QWR
size of  $a_x = a_y = 6$\ nm.
When $N_h = 4 \times 10^5$ m$^{-1}$, the SRT first increases 
then decreases with $T$ for the reason mentioned in the
paragraph above. When $N_h = 2 \times 10^6$ m$^{-1}$, one can see a
fast decrease of the SRT around $T=100$\ K. To understand this
behavior, we plot the hole distribution in the
$2\pm$  subbands in Fig.\ 6, from which one can see a fast
increase of population in the $2\pm$  subbands around $T=100$\
K, as the energy scale of the gap between the $1\pm$ and
$2\pm$ subbands is close to $k_B T$ of $T=100$\ K. This fast
increase of the hole occupation in the $2\pm$ subbands leads to an
increase of inter-subband spin-flip scattering which accounts for the
fast decrease of the SRT around $T=100$\ K. To further reveal the
contribution of inter-subband scattering, we plot the results which exclude
the inter-subband hole-phonon scattering and inter-subband hole-hole
scattering as  dashed curves in Fig.\ 6. One finds that the fast
decrease of the SRT around $T=100$\ K disappears.
We also plot the results without the coherent term but including
all the scattering as the chain curve. One can see that when $T<100$~K
and the holes are populated at the $1\pm$ subbands only, the chain curve
coincides with the solid curve for that the coherent term only includes
$\dot{\rho}^h_{k,1+,1-}|_{coh}$ which is negligible. When $T\ge
100$~K and holes populate both the $1\pm$ and $2\pm$ subbands,
there is difference between the chain curve and the solid curve
which is due to the inhomogeneous broadening in the coherent term
$\dot{\rho}^h_{k,1\pm,2\pm}|_{coh}$.  This inhomogeneous broadening,
together with the inter-subband scattering, can cause spin
relaxation as discussed 
in subsection A. However, the difference between the chain and
the solid curves is small. This indicates that the contribution of
this spin relaxation mechanism is not as important as the contribution
from the spin-flip scattering.

\subsection{Spin polarization dependence of SRT}

\begin{figure}
  \begin{center}
\includegraphics[width=9cm]{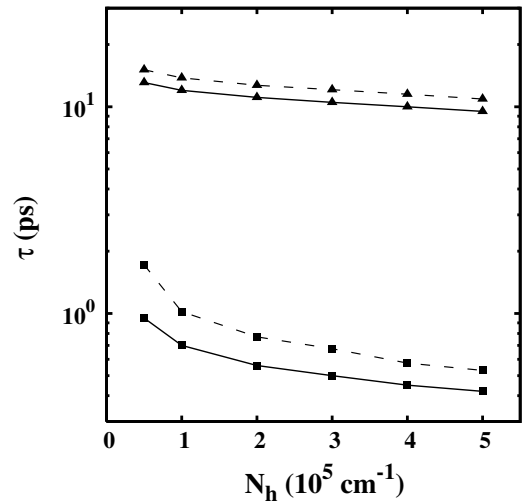}
  \end{center}
  \caption{SRT  {\em vs.} the hole density $N_h$ at different
    QWR sizes and initial spin polarizations.
    The solid  (dashed) curves are the results obtained for an initial
    spin polarization $P = 2.5 \%$ ($ 40 \%$). {\tiny
      $\blacksquare$}:  $a_x = a_y = 6$\ nm;
    {$\blacktriangle$}:  $a_x = a_y = 10$\ nm. $T = 100$~K.}
\end{figure}

Finally, we investigate the initial spin polarization dependence of the
spin relaxation. In Fig.\ 7 we plot the SRT as a function of hole
density for both low and high spin polarizations with different
QWR sizes at $T = 100$~K. It can be seen from the figure that
the SRT of the case with high spin polarization is larger. This originates
from the Hartree-Fock contribution of the hole-hole coulomb
interaction, which serves as an effective magnetic field 
and can effectively reduce the spin relaxation at large spin
polarization.\cite{Weng,schu}

\section{Conclusion}

In conclusion, we have investigated the spin relaxation of holes in
$p$-type GaAs QWRs. The SRT is calculated by
numerically solving the fully microscopic
kinetic spin Bloch equations in the helix spin space.
Differing from our previous works in $n$-type quantum-well and QWR
systems\cite{Wu1,Wu2,Wu3,Weng,Wu5,Cheng,Zhou} where the SOC is weak
and the collinear statistics is good enough, the helix statistics is
adopted in this investigation because of the strong SOC
for holes in QWR system. Using this approach, we
have studied in detail how the hole spin relaxation is affected by the wire
size, the hole densities, temperature and the spin polarization.
The confinement potential is assumed to be 
rectangular hard wall potential with
infinite-depth throughout the paper. In real sample, the SRT can be
 quantitatively different from our results due to the different
 confinements. However, the leading features such as the strong HH-LH
  mixing and the anticrossing points will be retained,\cite{Ulrich1}
  and the qualitative results will remain unchanged accordingly.

We show that when holes are populated at the lowest helix subbands
$1\pm$ only, the main spin-relaxation mechanism
is the spin-flip scattering which is proportional to the spin
mixing of the helix subbands. When holes are populated in both
$1\pm$ and $2\pm$ subbands, there are three mechanisms leading to
spin relaxation: first, the bare spin-flip scattering; second, the
spin-conserving scattering along with the inhomogeneous broadening;
and third, the spin-flip scattering  along with the inhomogeneous
broadening. However, the bare spin-flip scattering is still the
dominant spin relaxation mechanism.

The QWR size influences the SRT effectively because the spin
mixing and the subband structure in QWRs depend strongly on the
confinement.  When the wire width gets larger, the subbands are closer to
each other and the spin mixing of the subbands gets larger, therefore,
the SRT decreases.  Especially, in the case of $a_x=a_y=10$~nm, there
is an anticrossing point. If the Fermi surface
happens to be close to this point, this anticrossing point leads to a
strong spin-flip scattering which accounts for a minimum of the SRT.
When $a_x=20$~nm, $a_y=10$~nm, the anticrossing
point at small $k$ between the $1\pm$ and $2\pm$ disappears. However,
at large $k$, the lowest two subbands become very close to each
other. If we take the hole density to be $N_h = 2 \times
10^6$~cm$^{-1}$ to place the Fermi surface at this large $k$ region, the
SRT is nearly two orders of magnitude smaller than the SRT at small
wire width because the spin mixing here is very large and the
contribution from the spin-flip scattering is very strong.

The hole density influences the SRT by modulating the strength of
spin mixing and the strength of inter-subband spin-flip scattering. In
most of the cases we 
considered, the strength of spin mixing and the inter-subband
scattering increase with $N_h$ as the holes are populated in high $k$
states. As a result, the SRT decreases with $N_h$. However, when the
confinement is very strong and the energy gap between
$1\pm$ and $2\pm$ is large, there is a small region where the spin
mixing decreases with $k$. If we choose a small $N_h$ and a low
temperature to make the holes be distributed in this
small region only, one finds that the SRT increases with $N_h$ because
of the decreasing spin mixing.

The influence of temperature on the SRT is similar to the case of the
hole density dependence. The strength of both the spin mixing and the
spin-flip scattering increases with $T$, and the SRT decreases with
$T$. Especially, when the energy scale of the gap between the $1\pm$ and
$2\pm$ subbands is close to $k_B T$, there is a fast
increase of the distribution on the $2\pm$ subbands, which
leads to an
increase of inter-subband spin-flip scattering and leads to a
fast decrease of the SRT. This decrease of the SRT also proves that the
inter-subband spin-flip scattering makes a marked contribution to the
spin relaxation. We further show that the Hartree-Fock
term increases with spin polarization and can reduce the spin
relaxation.

%%%%%%%%%%%
\begin{acknowledgments}
This work was supported by the Natural Science Foundation of China
(Grants No.\ 10725417 and No.\ 10574120), the National
Basic Research Program of China (Grant
No.\ 2006CB922005), the Knowledge Innovation Project of the Chinese Academy
of Sciences, and the Royal Society of New Zealand (Grant No.\ ISATB06-62).
The authors acknowledge discussions with J. L. Cheng. One of
the authors (C.L.) thanks J. H. Jiang for valuable discussions. 
\end{acknowledgments}

\begin{appendix}

\section{A COMPARISON OF  BIA, SIA AND THE INTRINSIC
 SOC}

The SOC contributions for holes arising from the BIA and SIA can be
obtained by quasi-degenerate perturbation theory (L\"owdin
partitioning) from the extended Kane model.\cite{Winkler_book}
For an external electric field along $x$ direction, the dominant terms
of the SIA contribution can be written, in an explicit matrix notation, as
follows:\cite{Winkler_book}
\begin{equation}
  \label{SIA_matrix}
  H_{8v8v}^r = \frac{\gamma^{8v8v}_{41}{\cal E}_x}{\hbar}\left(
    \begin{array}{cccc} -\frac{3}{2}P_y & -\frac{\sqrt{3}}{2}iP_z & 0 &
      0 \\ \frac{\sqrt{3}}{2}iP_z &  -\frac{1}{2}P_y & -iP_z & 0 \\ 0 & iP_z &
      \frac{1}{2}P_y &  -\frac{\sqrt{3}}{2}iP_z \\ 0 & 0 &
      \frac{\sqrt{3}}{2}iP_z & \frac{3}{2}P_y  \end{array}
  \right).
\end{equation}
The coefficient $\gamma^{8v8v}_{41}$ for GaAs is $-1.462 \times
10^{-19}$ e$\cdot$m$^{2}$ and is two or three
orders of magnitude smaller than the off-diagonal terms in Eq.\
(\ref{Luttinger_Hamiltonian}).\cite{Winkler_book} 
Besides, this term couples the two LH states directly
while in Eq.\ (\ref{Luttinger_Hamiltonian}) the two LH
states can only mix with each other mediated by the HH
states. However, this direct coupling is still very small compared to
the intrinsic mixing due to the first term in Eq.\
(\ref{Luttinger_Hamiltonian}). To
show this, we study a simplified case including only the lowest eight
collinear subbands of $|1,1,\sigma\rangle$ and $|1,2,\sigma\rangle$
with $\sigma = \pm \frac{3}{2}$ and  $\sigma = \pm \frac{1}{2}$, and
we use the second-order L\"owdin partitioning to convert this
$8\times8$ matrix expanded by Eq.\ (\ref{Luttinger_Hamiltonian}) to a
block-diagonal form in which the off-diagonal matrix elements between
$|1,1,\pm\frac{1}{2}\rangle$ and the other states are zero. Then the
effective coupling between these two lowest LH states can be written
as:
\begin{eqnarray}
  \label{partitioning}
  H_{\frac{1}{2},-\frac{1}{2}}^{(2)} &=& \frac{\langle 1,1,\frac{1}{2}
  |S^{\dagger}|1,2,\frac{3}{2}\rangle \langle 1,2,\frac{3}{2}|R|
  1,1,-\frac{1}{2} \rangle}{E_{1,1,\frac{1}{2}} -
    E_{1,2,\frac{3}{2}}} \nonumber \\ &&\hspace{-1cm} +\frac{\langle
    1,1,\frac{1}{2} 
  |R|1,2,-\frac{3}{2}\rangle \langle 1,2,-\frac{3}{2}|-S^{\dagger}|
  1,1,-\frac{1}{2} \rangle}{E_{1,1,\frac{1}{2}} -
    E_{1,2,-\frac{3}{2}}} \nonumber \\ && = \frac{\hbar^2}{2 m_0}
  \frac{64 \gamma_2 \gamma_3 k}{a(3\gamma_1 - 13\gamma_2)},
\end{eqnarray}
in which we assume $a_x = a_y = a$ for simplicity. Then we compare
this term to the coupling between $|1,1,\pm\frac{1}{2}\rangle$
contributed by SIA, and find $\frac{\gamma^{8v8v}_{41}{\cal
    E}_xk}{H_{\frac{1}{2},-\frac{1}{2}}^{(2)}} = 4.1 \times 10^{-3}
$ when $a = 6 $\ nm and ${\cal E}_x = 100 $\ kV/cm (in experiments
with quantum wells, the values of ${\cal E}_x $ are typically of the
order of several kV/cm).\cite{Winkler2}  Therefore, the
contribution from SIA is very small.

The SOC contribution from the BIA is also very small. The BIA
coefficient $b^{8v8v}_{41}$ in Eq.\ (\ref{Dresselhaus}) is $-8.193
\times 10^{-29} $eV$\cdot$m$^3$ for GaAs,\cite{Winkler_book} and is one
order of magnitude smaller than the intrinsic 
mixing due to the first term in Eq.\ (\ref{Luttinger_Hamiltonian}).
Furthermore, when only the lowest four collinear states (two for HHs and two
for LHs) are included, one finds from Eq.\ (\ref{Dresselhaus}) that
only the third term
is nonzero. However, the third term is diagonal and does not
contribute any coupling. Therefore, the spin coupling due to the BIA
contributions must be mediated by higher collinear states.

\end{appendix}
%%%%%%%%%%%%%%%%%%%%%%%%%%%


\begin{thebibliography}{99}
\bibitem{comment} From our calculation, we find that
in the case of $a_x =a_y= 20$~nm,  $N_h = 2
\times 10^6$~cm$^{-1}$ and $T = 20$~K,  $\langle E \rangle$
is close to $3\pm$ states and the total population in $3\pm$ states is
approximately 8~\%.
\end{thebibliography}
\end{document}